\documentclass[preprint,aps]{revtex4}
\usepackage{graphicx}
\usepackage{dcolumn}
\usepackage{amsmath, amssymb}

\let\epsilon\varepsilon

\newcommand{\hatepsilon}{\myhat{\epsilon}}
\renewcommand{\hatepsilon}{{\boldsymbol \epsilon}}
\newcommand{\ie}{{\it i.e.}}
\newcommand{\comment}[1]{}

\begin{document}

\title{Variable Step Random Walks and Self-Similar Distributions}

\author{Gemunu H. Gunaratne,$^{1,2}$ Joseph L. McCauley,$^{1}$ Matthew Nicol,$^{3}$
and Andrei T\"{o}r\"{o}k$^{3,4}$}

\address{$^{1}$ Department of Physics,
                University of Houston,
                Houston, TX 77204}
\address{$^{2}$ Institute for Fundamental Studies,
                Hantana, Sri Lanka}
\address{$^{3}$ Department of Mathematics,
                University of Houston,
                Houston, TX 77204}
\address{$^{4}$ Institute of Mathematics of the Romanian Academy, 
                Bucharest, Romania}
\nobreak

\begin{abstract}
  We study a scenario under which variable step random walks give anomalous
  statistics. We begin by analyzing the Martingale Central Limit Theorem to
  find a sufficient condition for the limit distribution to be
  non-Gaussian.  We note that the theorem implies that the scaling
  index~$\zeta$ is~$\frac{1}{2}$. For corresponding continuous time
  processes, it is shown that the probability density function $W(x;t)$
  satisfies the Fokker-Planck equation. Possible forms for the diffusion
  coefficient are given, and related to $W(x,t)$.  Finally, we show how a
  time-series can be used to distinguish between these variable diffusion
  processes and L\'evy dynamics.
\end{abstract}
\pacs{PACS number(s): 05.40.Fb, 05.40.-a, 47.27.-1, 05.10.Gg, 89.65.Gh}
\maketitle

\section {Introduction}\label{sec.introduction}
Under which conditions can statistics of stochastic processes be anomalous?
Such statistics have been observed in
temperature and longitudinal velocity fluctuations in highly turbulent
fluid flows \cite{hesAcas,casAgun,wuAkad,solAgol,takAseg}, instantaneous
velocities of gusting winds \cite{embAklu} and price variations in
financial markets \cite{mand,manAsta,friApei,arnAmuz,dacAgen,macAgun}.
Furthermore, in the case of financial markets, it has been noted that the
probability density functions exhibit self-similarity \cite{mand,manAsta}.
In prior work, L\'evy statistics \cite{klaAsch,bouAgeo,solAwee} and hierarchical
processes \cite{peiAbot} have been suggested as possible causes of 
anomalous behavior. In this paper we suggest an alternative scenario where,
in contrast to L\'evi processes, stochastic increments have uniformly bounded 
variance and are not independent.

General conditions for the validity of the Central Limit Theorem (CLT) are
given for \emph{martingales}, which are defined in Section~\ref{sec.MCLT}.
We discuss how the limit distribution can fail to be Gaussian, and provide a 
set of conditions that give non-Gaussian statistics.  In Section~\ref{sec.markov}, 
we argue that the corresponding probability density $W(x;t)$ for continuous 
time processes satisfies the Fokker-Planck equation \cite{fokApla,chan} and 
scales with index $\zeta (=\frac{1}{2})$; \ie, $W(x;t) = \frac{1}{\sqrt{t}} F(u)$,
where $u=\frac{x}{\sqrt{t}}$. By using the Fokker-Planck equation, we show
that the diffusion coefficient $D(x;t)$ for the process takes a specific
form.  It is shown that a reduction of $D(x;t)$ to a form $D(u)$ preserves
all statistical features of the stochastic process. Given $D(u)$, we then
provide an explicit expression for $F(u)$.  As examples, we provide forms
for $D(u)$ that give exponential and power-law distributions for $F(u)$. In
Section~\ref{sec.distinguishing}, we provide a criterion that can be used
to distinguish these newly introduced variable diffusion processes from
L\'evy statistics. Previous analysis of fluctuations in financial markets
appear to contradict the L\'evy mechanism.

Throughout the paper, we will relate our results to turbulent flows and
financial markets to illustrate implications of our assertions. However, it
should be emphasized that our work is a theory of neither of these systems.

\section {Martingale Central Limit Theorem}\label{sec.MCLT}

The classical CLT in the context of identical independently distributed
events $\{\epsilon_k\}$ with zero mean and variance $\sigma^2$, states that
\begin{equation}
\frac{1}{\sqrt{n}} \sum_{k=1}^{n} \epsilon_k \to {\cal N} (0, \sigma^2)
\label{clt}
\end{equation}
as $n \rightarrow \infty$, where ${\cal N}(0,\sigma^2)$ denotes a zero-mean
normal distribution. Here the convergence is \emph{in distribution}; \ie,
for each $a$,
\begin{equation}
\lim_{n\to\infty}P\left(\frac{1}{\sqrt{n}} \sum_{k=1}^{n} \epsilon_k \le a
\right) = \frac{1}{\sqrt{2\pi}\sigma}\int_{-\infty}^a e^{-x^2/2\sigma^2} d
x.
\label{conv_distn}
\end{equation}
The CLT can be generalized for a class of non-independent processes,
referred to as {\it martingales}. We describe the setup only for
the case of interest to us.

We consider a random walk (on the real line) starting at $x_0=0$, with
steps denoted by $\hatepsilon = (\epsilon_1, \epsilon_2, \dots)$.  The step
$\epsilon_k$ can depend on its history (\ie, the previous $(k-1)$ steps).
The position after $n$ steps is denoted by $x_n=\sum_{k=1}^{n} \epsilon_k$.
A probability measure $p$ is given on the space of the infinite sequences;
$p$ induces a measure $p_n(\hatepsilon^{(n)})$ on the space of first $n$
steps $\hatepsilon^{(n)}$. When the context is clear, we denote $p_n$ by
$p$. If the random variables $\epsilon_k$ are discrete, then $p_n$ includes
$\delta$-functions.

The conditional probability of the $k^{th}$ step given the history
$\hatepsilon^{(k-1)}$ is defined by
\begin{equation}
p\left(\epsilon_k | \hatepsilon^{(k-1)} \right) =
\frac{p\left(\hatepsilon^{(k)} \right)}{p\left(\hatepsilon^{(k-1)}
  \right)}.
\label{conditionalp}
\end{equation}
The random variables $\{x_n\}_{n\ge 0}$ form a martingale if each increment
(or \emph{martingale difference}) $\epsilon_k$ has zero conditional mean:
\ie, if for each $k\ge 1$
\begin{equation}
E \left[\epsilon_{k} | \hatepsilon^{(k-1)}\right] = 0,
\label{martingale}
\end{equation}
for all histories $\hatepsilon^{(k-1)}$. (Note that, although this mean
value is independent of the history of the walk, the conditional
probability density given in Eqn. (\ref{conditionalp}) can depend on 
$\hatepsilon^{(k-1)}$.)
Finally define the expected value of the location over all $n$-step random
walks by
\begin{equation}
E\left[x_n\right] = \int  \ x_n \
d p\left(\hatepsilon^{(n)} \right) = \int d\hatepsilon^{(n)} \ x_n \
p\left(\hatepsilon^{(n)} \right), 
\label{expectation}
\end{equation}
where the second formula is written just to emphasize the variables over
which the integration takes place. Denote the corresponding variance by
$Var\left[x_n\right]$. We have the following Lemma for martingale processes. 

\noindent {\bf Lemma:} If $\{x_n\}$ is a martingale process with
$x_0\equiv 0$, then
\begin{enumerate}
\parskip = -0.5pt
\itemsep = -0.5pt
\item [(I)] $E\left[ x_n \right] = 0$.
\item [(II)] $Var \left[ x_n \right] = \sum_{k=1}^{n} Var \left[ \epsilon_k
  \right]$.
\end{enumerate}

These results can be proved inductively using \footnote{These properties 
  actually follow from $E[x_n]=E[x_{n-1}]$, and $E\left[x_n \epsilon_{n+1} \right] =
  E\left[x_n E\left[\epsilon_{n+1} | \hatepsilon^{(n)} \right] \right]=0$,
  for any martingale.}
\begin{equation}
   \int d \epsilon_{n+1} p \left(\epsilon_{n+1} | \hatepsilon^{(n)}
   \right) = 1,
\label{condition1}
\end{equation}
and
\begin{equation}
  \int d \epsilon_{n+1} \epsilon_{n+1} p\left(\epsilon_{n+1} |
  \hatepsilon^{(n)} \right) = E\left[\epsilon_{n+1} |
  \hatepsilon^{(n)}\right] = 0.
\label {condition2}
\end{equation}

For martingales, Theorem 3.2 of Ref. \cite{Hall-Heyde} gives a
more general form of the CLT.
Recall that a sequence $y_n$ of random variables is said to converge
\emph{in probability} to a random variable $y$ if for any $\delta >0$, the
probability of $|y_n-y| > \delta$ goes to zero as $n\to\infty$.

\noindent {\bf Theorem} (Martingale Central Limit Theorem)
Suppose that $\epsilon_1, \epsilon_2,\ldots$ are square-integrable
martingale differences such that
\begin{itemize}
\parskip = -0.5pt
\itemsep = -0.5pt
\item[(1)] $\max_{1\le k \le n} \left(|\epsilon_k|/\sqrt{n}\right)
  \rightarrow 0$ in probability,
\item[(2)] $\sum_{k=1}^{n}\epsilon_k^2/ n  \rightarrow \eta^2$ in probability,
\item[(3)] $ E\left[ \max_{1\le k \le n} \left(\epsilon_k^2 / n\right)
  \right]$ is bounded in $n$,
\end{itemize}
where the random variable $\eta$ is finite with measure 1. Then
\begin{equation}
\frac{1}{\sqrt{n}} \sum_{k=1}^{n}\epsilon_k \rightarrow Z,
\label{martingaleclt}
\end{equation}
where the convergence is in distribution (see Eqn. (\ref{conv_distn})), and
the random variable $Z$ has characteristic function (\ie,
$E\left[exp(itZ)\right]$) given by
\begin{equation}
  \label{eqn.char-function}
  E\left[exp(itZ)\right]=E\left[exp(-\frac{1}{2}\eta^2 t^2)\right]\quad
  \text{ for all } t.
\end{equation}

Observe that the martingale differences $\epsilon_k$ are not required to
be independent or to be distributed identically. However, when the conditions 
of the theorem are satisfied, the distribution of the random variable 
$u_n=\frac{x_n}{\sqrt{n}}$ converges to $F(u)$, the distribution of $Z$. 
We will refer to this property as \emph{scalability with scaling index $\zeta=\frac12$}.

We first provide a necessary and sufficient condition to obtain Gaussian 
statistics.

\noindent {\bf Lemma:} If the random variables $Z$ and $\eta$ satisfy
Eqn. (\ref{eqn.char-function}), then $Z$ is Gaussian if and only if
$\eta^2$ is a constant.

Indeed, if $\eta^2$ is constant, say $\sigma^2$, then $Z$ has
characteristic function $exp\left(-\frac{\sigma^2 t^2}{2}\right)$, and
therefore it is normally distributed with variance $\sigma^2$. Conversely,
if $Z$ has mean zero and is a Gaussian with variance $\sigma^2$, then its
characteristic function equals $exp\left(-t^2 \sigma^2/2\right)$. Write
$s=\frac{1}{2} t^2$, $g=\eta^2$, and define the probability measure
$d\widetilde{P}=e^{\sigma^2}e^{-g}dP$. Then all moments of $g$ exist with
respect to $\widetilde{P}$. The equality
\[
e^{-\sigma^2 s}=\int e^{-g s} d {P}
\]
implies, upon differentiating with respect to $s$ and setting $s=1$, that
$\widetilde{E}[g^n]=\sigma^{2n}$ for all $n$. Thus $\eta^2$ is constant
(Theorem 3.11 of Ref. \cite{Durrett}).

Next, we identify a set of conditions that gives anomalous statistics for
$u_n$. Condition (3) of the martingale CLT is satisfied if increments
$\epsilon_k$ have a variance bounded uniformly in $n$ (\ie, there exists a
$c>0$ such that for all $k$, $Var\left[\epsilon_k\right] =
E\left[\epsilon_k^2\right] \le c$). To see this, note that
\[
E\left[ \max_{1\le k \le n} \left(\epsilon_k^2 / n\right) \right] \le 
  E\left[\sum_{k=1}^{n} \epsilon_k^2 / n\right] 
  = \frac{1}{n} \sum_{k=1}^n Var \left[ \epsilon_k \right] \le c,
\]
where the equality follows from the previous Lemma. Condition (1) is
satisfied under the stronger property that there exists $\delta > 0$ and
$c_1>0$ such that for all $k$
\begin{equation}
  E[|\epsilon_k|^{2+\delta}] \le c_1.
\label{stronger}
\end{equation}
This can be seen from 
\[
Prob\left(\max_{1\le k \le n} \left(|\epsilon_k|/\sqrt{n}\right)>\beta
\right)\le \sum_{k=1}^n Prob\left(|\epsilon_k|/\sqrt{n}>\beta\right)
=\sum_{k=1}^n Prob\left(|\epsilon_k| > \sqrt{n}\beta\right)
\]
and the fact that
\[
c_1 \ge E[|\epsilon_k|^{2+\delta}] \ge (\sqrt{n}\beta)^{2+\delta}
Prob\left(|\epsilon_k| > \sqrt{n}\beta\right).
\]
Therefore $Prob\left(|\epsilon_k| > \sqrt{n}\beta\right)\le
{c_1}/{\sqrt{n}^{2+\beta}}$, which implies that
\[
Prob\left(\max_{1\le k \le n} \left(|\epsilon_k|/\sqrt{n}\right)>\beta
\right) \le n \frac{c_1}{\sqrt{n}^{2+\beta}} \to 0 \quad \text{ as }n\to
\infty.
\]

What remains is to find martingales that satisfy condition (2) where
$\eta^2$ is not a constant. If $\epsilon_k$'s are independent and
identically distributed with finite variance $\sigma^2$, then from the
classical CLT $\eta^2 = \sigma^2$ in probability. Once $\epsilon_k$'s are
allowed to be history dependent, the conditions for convergence of
$\left(\sum \epsilon_k^2 / n\right)$ becomes non-trivial, as illustrated by
the following example: consider a stochastic process with independent
steps, consisting of $N_1$
steps from a distribution with finite variance $\sigma_1^2$, followed by
$M_1$ steps from a distribution with finite variance $\sigma_2^2$, followed
by $N_2$ steps from the first process, $M_2$ steps from the second, etc.
For suitable choices of $N_1 \ll M_1 \ll N_2 \ll M_2 \ll \ldots$, $\eta^2$
moves between $\sigma_1^2$ and $\sigma_2^2$, and fails to converge.
Convergence of $\eta^2$ requires more stringent conditions on the
stochastic process.

For processes introduced in the next section, the distribution of $\eta^2$
is not a $\delta$-function, as shown in the Appendix.

We conclude this section with the following observations. First, we
reiterate that once the conditions of the martingale CLT are satisfied,
$u_n = x_n/\sqrt{n}$ converges to a distribution $F(u)$; i.e., the scaling
index $\zeta$ is $\frac{1}{2}$. Second, in contrast to L\'evy processes,
increments $\epsilon_k$ are not independent. Further, for the examples we
consider, the $\epsilon_k$'s satisfy Eqn. (\ref{stronger}) (at least,
according to the numerical simulations). Note however that the conditional
variance, $Var \left[ \epsilon_{n+1} | \hatepsilon^{(n)} \right]$, is not
required to be uniformly bounded.

\section{Continuous Markov Processes}\label{sec.markov}

In order to study continuous processes, divide the interval $t$ into
subintervals of $\delta t$ and let $n=t/\delta t$; it is assumed that
$\delta t$ is sufficiently large for many martingale increments to occur
in this interval. Now,
let $\epsilon_k$'s denote the martingale increments in intervals $\delta
t$. In order for the variance of increments in 1 unit of time to be
uniformly bounded, it is necessary and sufficient (see the first Lemma) that
$\frac{1}{\delta t} Var\left[\epsilon_k\right]$ be uniformly
bounded; \ie, that $Var\left[\epsilon_k/\sqrt{\delta t}\right]$ be
uniformly bounded. A priori, the limit $Z$ may depend on the particular
discretization used. For the examples given below, this is not the case,
although we have not been able to derive it analytically as yet.

For the remainder of the paper, we limit considerations to Markov
processes; \ie, $ p \left( \epsilon_k | \hatepsilon^{(k-1)} \right) = p
\left( \epsilon_k | x_{k-1}; (k-1) \right)$ for each $k$; here, the
possible dependence of the probability density on the step number (see,
Section~\ref{sec.MCLT}) is denoted explicitly. Markov processes satisfy the
master equation \cite{chan}
\begin{equation}
W\left(x; t+\delta t \right) = \int d\epsilon W\left( x-\epsilon; t\right)  p_{\delta t} \left(\epsilon | (x-\epsilon); t\right),
\label{mastereqn}
\end{equation}
where $p_{\delta t} \left(\epsilon | x; t \right)$ denotes the probability
density function for an increment $\epsilon$ to occur in time $\delta t$
beginning from $(x; t)$. Taylor expanding in the variables $t$ and $x$
about $W(x; t)$, and noting that $Var \left[\epsilon / \sqrt{\delta
    t}\right]$ is bounded, gives the Fokker-Planck equation
\cite{fokApla,chan,masAzha}
\begin{equation}
  \frac {\partial}{\partial t} W(x;t)= \frac{1}{2} \frac {\partial^2
  }{\partial t^2} \left( D(x;t) W(x;t) \right),
\label {FPeqn}
\end{equation}
where the \emph{diffusion coefficient} $D(x;t)$ is given by
\begin{equation}
D(x;t) = \frac{1}{\delta t} \int d\epsilon \ \epsilon^2 p \left(\epsilon |
  x; t \right) = Var \left[ \frac{\epsilon}{\sqrt{\delta t}} | x;t \right].
\label{diff_coef}
\end{equation}
The derivation assumes the martingale condition $E\left[ \epsilon | x; t
\right] = \int d\epsilon \ \epsilon \ p_{\delta t} (\epsilon | x; t) = 0$.

Observe next that, since the scaling index $\zeta = \frac{1}{2}$, the
probability density for scalable martingales can be written as
\begin{equation}
    W(x;t) = \frac{1}{\sqrt{t}} F(u),
\label{scaling}
\end{equation}
where $u= x / \sqrt{t}$, and the pre-factor $1/\sqrt{t}$ has been included
in order that $W(x;t)$ be normalized (\ie, $\int dx W(x;t)$ be
time-independent). Only certain forms of $D(x;t)$ can be consistent with
this requirement. In order to obtain them, change variables so that $D(x;t)
= \bar D (u;t)$. Substituting in the Fokker-Planck equation gives
\begin{equation}
\frac{\partial}{\partial u} \left( u F(u) \right) +
\frac{\partial^2}{\partial u^2} \left(\bar D(u;t) F(u) \right) = 0,
\label{FPu}
\end{equation}
which can be integrated to
\begin{equation}
 u F(u) + \frac{\partial}{\partial u} \left( \bar D(u;t) F(u) \right) = c_1(t).
\label{firstint}
\end{equation}
Here $c_1(t)$ is the ``constant'' of integration. Integrating a second time
gives
\begin{equation}
\bar D (u;t) = -\frac{1}{F(u)} \int_{-\infty}^{u} dv \ v F(v) +
\frac{1}{F(u)} \left(c_1(t) u + c_2 (t) \right).
\label{secondint}
\end{equation}
where $c_2(t)$ is the second constant of integration.

\noindent {\bf Examples:} The Gaussian distribution $F(u) = exp
\left(-\frac{1}{2} u^2 \right)$ corresponds to $\bar D (u;t) = 1 + \left(u
  c_1(t) + c_2(t)\right) exp \left( \frac{1}{2}u^2 \right)$. The
exponential distribution $F(u) = exp \left( -|u| \right)$ corresponds to a
diffusion coefficient $\bar D (u;t) = (1+|u|) + \left(u c_1(t) +
  c_2(t)\right) exp \left( -|u| \right)$.

Note that the terms in $\bar D(u;t)$ that contain $c_1(t)$ and $c_2(t)$ do
not change the form of $W(x;t)$. Hence, they will not be considered in the
remainder of the paper; \ie, only the $t$-independent part of $\bar D
(u;t)$, henceforth denoted $D(u)$, will be considered.

Conversely, if the diffusion coefficient $D(u)$ is given, Eqn. (\ref{FPu})
can be integrated to give
\begin{equation}
F(u) = \frac{1}{D(u)} exp \left(-\int^{u} dv \frac{v}{D(v)} \right) \left[
  a_1 \int^{u} dv \ exp \left( \int^{v} dw \frac{w}{D(w)} \right) + a_2
\right],
\label {fsoln}
\end{equation}
where $a_1$ and $a_2$ are constants of integration. If $D(u)$ is symmetric
under reflections about the origin and the process begins at $x_0=0$ then
$F(u)$ is symmetric \footnote{It is possible that the symmetric $F(u)$ is
  unstable, and the stable distributions consists of a pair of functions
  related by reflectional symmetry. In the examples given here, the
  symmetric $F(u)$ is found to be the solution of the Langevin equation for
  motion starting from the origin.}; consequently $a_1=0$, as can be seen
from the anti-symmetry of the left side of Eqn. (\ref{firstint}). Then,
\begin{equation}
F(u) = \frac{1}{D(u)} exp \left(-\int^{u} dv \frac{v}{D(v)} \right).
\label{fsoln2}
\end{equation}
The form of $F(u)$ for selected diffusion rates is given next.  As
mentioned in Section~\ref{sec.MCLT}, although $Var \left[\epsilon_n /
  \sqrt{\delta t}\right] $ for each $n$ is uniformly bounded, the conditional
variance $Var \left[\epsilon_n / \sqrt{\delta t}| x;t\right]$ of the
martingale differences, given by $D(u)$, is not required to be bounded with
respect to $u$.

\noindent{\bf Examples:}
\begin{itemize}
\parskip = -0.5pt
\itemsep = -0.5pt
\item [(I)] $D(u) = 1 \longrightarrow F(u) = exp \left( -\frac{1}{2} u^2
  \right)$
\item [(II)] $D(u) = 1 + \alpha |u| \longrightarrow F(u) = exp \left(
    -\frac{|u|}{\alpha} \right) / \left(1 + \alpha |u|
  \right)^{(1-\alpha^{-2})} $
\item [(III)] $ D(u) = 1 + |u| \longrightarrow F(u) = exp \left(-|u|
  \right) $
\item [(IV)] $ D(u) = \left( 1+\alpha u^2 \right) \longrightarrow F(u) =
  \left(1 + \alpha u^2 \right)^{-\left(1 + (1/2\alpha)\right)}$
\end{itemize}
Thus, suitable choices of $D(u)$ can give exponential or power-law behavior
in $F(u)$. Note that, in the final example $\alpha < 1$ is needed in order
for the condition (\ref{stronger}) to be satisfied.

We have confirmed numerically that stochastic dynamics with diffusion
coefficients given in these examples give probability density functions
consistent with the analytically derived expressions. These computations
were conducted by integrating the (zero-drift) Langevin equation $dX =
[D(X(t);t)]^{1/2} {\cal N}(0,dt)$~\cite{gill}. The integrations are done
using Ito calculus; \ie, it is assumed that each step in the integration
consists of a large number of stochastic increments and that variations in
$D(X(t); t)$ during the interval can be ignored. Consequently, the
deviations in a time interval $\delta t$ lie on ${\cal N} (0, \sqrt{D}
\delta t)$.

\section {Distinguishing Between L\'evy and Variable Diffusion
  Processes}\label{sec.distinguishing}

\begin{figure}
\includegraphics[width=3.0in]{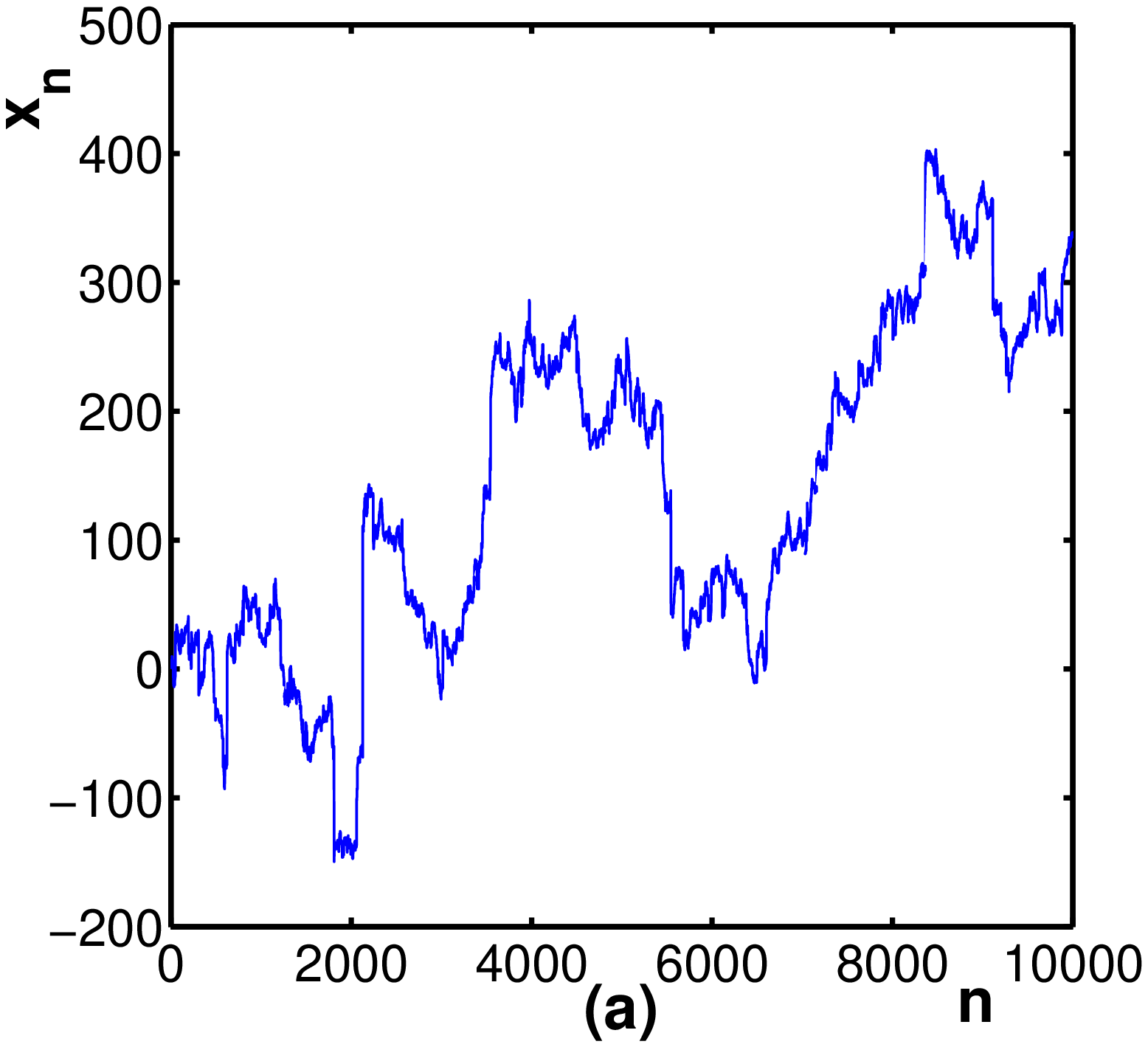}
\includegraphics[width=3.0in]{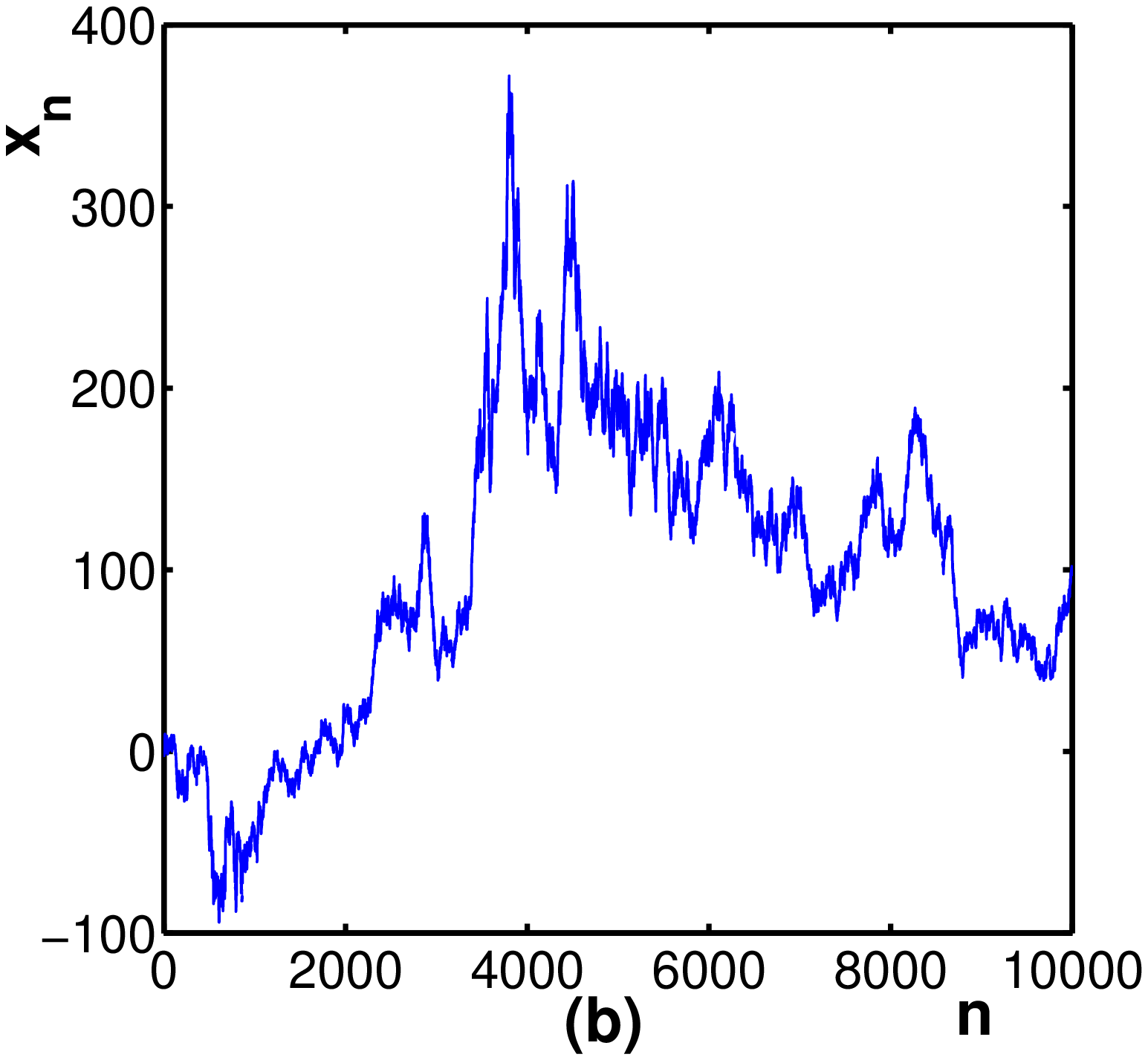}
\caption {Examples of 10,000 steps from a (a) L\'evy distribution with $\zeta=2/3$ and 
  (b) variable diffusion process with $D(u) = 2(1+u^2)$.  Unlike in (a), a
  large fluctuation in (b) is generally followed by movements with higher
  amplitude.}
\label{runs}
\end{figure}

Given a stochastic process $\{\epsilon_k\}$ such that $x_n$ lies on a
scalable, non-normal distribution, is it possible to determine if L\'evy or
variable diffusion processes are the more likely source of the dynamics?
More precisely, is it possible to eliminate one of the scenarios as the
underlying cause of the observed stochastic dynamics? One possible
criterion is to test if the variance of the stochastic process is finite
(variable diffusion) or infinite (L\'evy). However, it is difficult to make
this determination from a finite time series. An alternative is to use the
fact that while successive movements of a L\'evy process are independent,
those in the variable diffusion case depend on the location and time of the
walk. For example, if $D(u)$ increases with $|u|$ (as in the examples
above), then large movements are likely to leave $x_n$ (and hence $D(u)$)
large; consequently, a large fluctuation can generally be expected to be
followed by additional (positive or negative) large increments. L\'evy
processes with independent increments will not exhibit such correlations.
Dynamics of L\'evy and variable diffusion processes, shown in Figure
\ref{runs}, illustrates the difference.

Thus, one may consider distinguishing variable diffusion and L\'evy
processes using the auto-correlation function of a time series.  However,
since the mean value of the increments is zero in for each case (since they
are martingales), the auto-correlation will vanish. On the other hand,
auto-correlation function of $\{\epsilon_k^2\}^{(n)}$ will only vanish for
the L\'evy case. Specifically, for a random time series of length $n$, we
use
\begin{equation}
{\cal C}(m;n) \equiv \frac{1}{Var [\epsilon^2]} \left\langle
             \left(\epsilon^2_{k} - \langle \epsilon^2\rangle\right)
             \left(\epsilon^2_{k+m} - \langle \epsilon^2 \rangle \right)
             \right\rangle,
\label{correlation}
\end{equation}
where $\langle . \rangle$ denotes the average over $k$.  For L\'evy
processes, ${\cal C}(m;n)$ vanishes for $m>0$, while for variable diffusion
processes with $D(u) = 1 + |u|$, it is found to decay as $exp (-\alpha m /
n)$; the $n$-dependence implies that a longer series contains larger
fluctuations.

For fluctuations in financial markets, ${\cal C}(m;n)$ is known to exhibit
a slow decay with $m$ \cite{conAbou}. This phenomenon, referred to as
``clustering of volatility,'' suggests that L\'evy processes are unlikely to
be the source of scalable non-Gaussian distributions in financial markets.

\section{Discussion}\label{sec.discussion}

The theory we have presented is not merely a reformulation where an
observed scalable probability density function $W(x;t)$ is recast into a
suitably chosen diffusion coefficient $D(x;t)$. Rather, it introduces a new
class of stochastic dynamics. Unlike L\'evy processes, the increments
considered in our work, although Markovian, are not independent. In
addition, they have finite variances. The scaling index for scalable
diffusion processes takes a unique value $\zeta = \frac{1}{2}$. The
probability density function $W(x;t)$ for continuous time stochastic
dynamics takes the form $\frac{1}{\sqrt{t}} F(u)$ and satisfies the
Fokker-Planck equation. The diffusion coefficient can be chosen to be a
function of $u$, and there is a correspondence between $F(u)$ and the
diffusion coefficient $D(u)$.

The fact that successive events are independent in L\'evy processes and
only martingales in our variable diffusion processes implies that dynamics
can be used to identify which model is more suitable to represent a given
time series of stochastic events. We propose the use of the
auto-correlation of $\epsilon_k^2$'s as such a test. Previous studies of
financial markets suggest that they consist of increments that are not
independent, and hence suggest that independent L\'evy processes are
unlikely to be the correct explanation for the observed non-Gaussian
probability density functions \cite{conAbou}.

The need for $x$-dependent diffusion coefficients implies that the
stochastic dynamics is not invariant under translations in $x$. In
particular, for the examples given earlier, the origin is both the starting
point of the walk as well as the location where $D(x;t)$ is minimized. In
financial markets, one does expect any sudden large fluctuation in the
price of a stock to be followed by a period of high anxiety in the part of
traders; consequently the stock can be expected to trade at a significantly
higher rate. This is equivalent to an increase in the diffusion rate.
However, if the price of the stock settles at this new value, it is likely
that the location of the minimum in $D(x;t)$ will move towards it. Thus, a
more realistic model of financial markets would involve a coupled variation
of the price of the stock and the location of the minimum of the diffusion
coefficient \cite{aleAbas}.

A time-dependent, but $x$-independent drift $\mu(t)$ of the stochastic
process can be introduced by including a ``drift" term $-\mu(t) W(x;t)$ on
the right side of the Fokker-Planck equation \cite{chan}. Redefining $u$ to
be $\frac{1}{\sqrt{t}} \left(x - \int^{t} \mu(s) ds\right)$ gives Eqn.
(\ref{FPeqn}), and the rest of the analysis presented here follows.

\section {Acknowledgements}
The research of GHG is partially supported by the NSF Grant 
PHY-0201001 and a grant from the Institute of Space Science Operations 
at the University of Houston (GHG).
The research of M. Nicol and A. T\"or\"ok was supported in part by NSF
Grant DMS-0244529.

\emph{It is a great pleasure to dedicate this paper to Mitchell Feigenbaum
  on the occasion of his $60^{th}$ birthday. Mitchell's outlook on Science,
  Arts, and Philosophy have been a source of inspiration for GHG for over
  20 years.}

\appendix

\section {Anomalous Martingale Processes}
\label{sec.appendix}

When the diffusion coefficient is a function of $u$, the martingale sums
may fail to lie on a normal distribution. We have chosen processes where
$E\left[|\epsilon_k|^{2+\delta}\right]$ is uniformly bounded, so that 
conditions (1) and (3) of the martingale CLT are satisfied. Hence, the 
random variable $Z$ is not distributed normally because 
$\left(1/n\right) \sum \epsilon_k^2 $ does not
approach a constant (in probability) for large $n$. We illustrate this
failure with two examples of discrete random walks.

\begin{figure}
\includegraphics[width=3.0in]{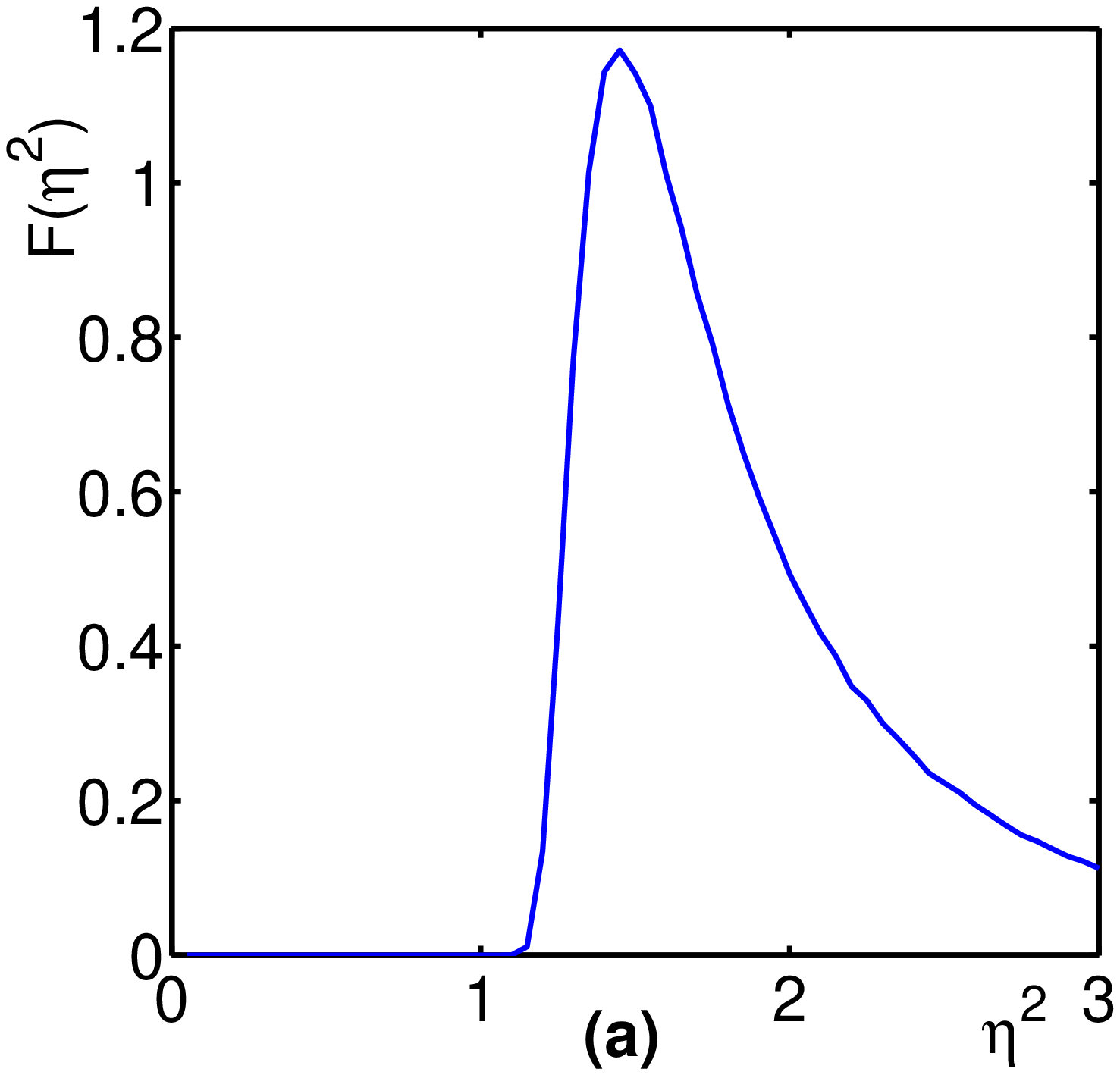}
\includegraphics[width=3.0in]{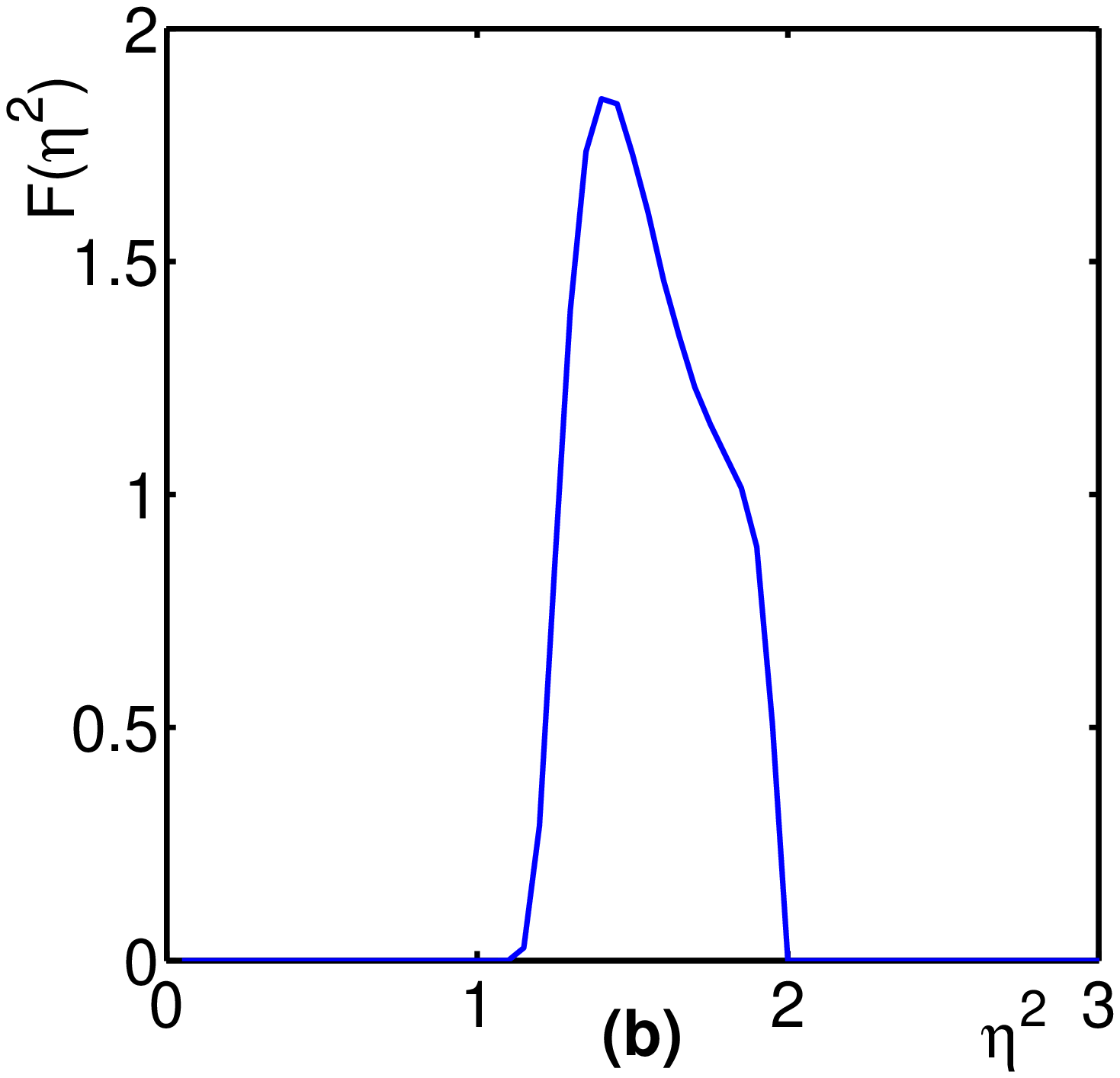}
\caption {The density function $F$ of $\eta^2 = \lim_{n\to\infty} \frac1n
  \sum_1^n\epsilon_k^2$ for random walks with (a) $D(u) = 1+|u|$, and (b)
  $D(u) = 1 + tanh(|u|)$. The fact that they are not $\delta$-functions
  implies that $\lim \left(x_n / \sqrt{n} \right)$ is not Gaussian, see
  Section~\ref{sec.MCLT}.}
\label{eta}
\end{figure}

The distribution of $\eta^2$ for a finite-step martingale with $D(u) =
1+|u|$ is shown in Figure \ref{eta}(a). Since $D(u) \ge 1$ for all $u$,
$\eta^2$ is non-vanishing only when the argument is larger than 1, where it
decays exponentially. As expected from the analysis, $F(u)$ is found to be
$\frac{1}{2} exp \left( -|u| \right)$.

Next, consider a martingale with $D(u) = \left(1 + tanh\ |u|\right)$. For a
fixed $t$, $D(u)$ varies between 1 and 2, and for a fixed $x$, it reduces
to 1 with increasing $t$. The histogram of $\eta^2$, computed numerically for
a set of 100,000 random walks of length 100,000, converges to the function
shown in Figure \ref{eta}(b). Since $1 \le D(u) \le 2$, $\eta^2$ is
non-zero only in the interval $[1,2]$. The corresponding probability
density function $W(x;t)$ has the form $\frac{1}{\sqrt{t}} F(u)$, but
$F(u)$ is not Gaussian. In contrast, if the diffusion coefficient is chosen
to be $\left(1 + tanh\ |x|\right)$ or $\left( 1 + tanh\ (1/\sqrt{t})
\right)$, $\eta^2$ is found to be constant, and $W(x;t)$ is found to
approach a Gaussian.

\end{document}